\documentclass{iaus}
\usepackage{graphicx}

\title[jet launching] 
{The role of thermal pressure in jet launching}

\def \cm{~\rm{cm}}
\def \s{~\rm{s}}
\def \km{~\rm{km}}

\def \g{~\rm{g}}

\def \AU{~\rm{AU}}

\def \yr{~\rm{yr}}

\author[Noam Soker]   
{Noam Soker$^1$ } %

\affiliation{$^1$Department of Physics, Technion, Haifa 32000, Israel }

\editors{J. Bouvier, eds.}
\begin{document}

\maketitle

\begin{abstract}

I present and discuss a unified scheme for jet launching that is based
on stochastic dissipation of the accretion disk kinetic energy, mainly
via shock waves.
In this scheme, termed thermally-launched jet model, the kinetic energy
of the accreted mass is transferred to internal energy, e.g., heat or
magnetic energy.
The internal energy accelerates a small fraction of the accreted mass to high
speeds and form jets.
For example, thermal energy forms a pressure gradient that accelerates the gas.
A second acceleration stage is possible wherein the primary outflow stretches magnetic
field lines. The field lines then reconnect and accelerate small amount of mass
to very high speeds. This double-stage acceleration process might form highly relativistic
jets from black holes and neutron stars.
The model predicts that detail analysis of accreting brown dwarfs that launch jets
will show the mass accretion rate to be
$\dot M_{\rm BD} \gtrsim 10^{-9}-10^{-8} M_\odot \yr^{-1}$,
which is higher than present claims in the literature.
\end{abstract}

\firstsection 

\section{Introduction}
In many popular models for the formation of astrophysical massive jets
(to distinguish from low density hot-plasma jets from radio pulsars)
magnetic fields play a dominate role in accelerating the jet's
material from the accretion disk.
Most models in young stellar objects (YSOs) are based on the
operation of large scale magnetic fields driving the flow from
the disk.
In the ``X-wind mechanism'' introduced by Shu et al. (1988, 1991)
the jets are launched from a narrow region in the magnetopause of the
stellar field.
A different model, although also uses open radial magnetic field lines,
is based on an outflow from an extended disk region,
and is not rely on the stellar magnetic field
(Ferreira \& Pelletier 1993, 1995; Wardle \& K\"onigl 1993;
K\"onigl \& Pudritz 2000; Shu et al. 2000; Ferreira 2002; Krasnopolsky et al. 2003;
Ferreira \& Casse 2004).

Other MHD simulations show that the high post-shock
thermal pressure might accelerate gas and form jets and/or
winds, e.g., as in the accretion around a black hole (BH) simulations
performed by De Villiers et al. (2004; also Hawley \& Balbus 2002).
In simulations of accretion onto a rotating BH De Villiers et al.\ (2005)
find that both gas pressure gradients and Lorentz forces in the inner
torus play a significant role in launching the jets.

A confining external pressure is required at the edge of accelerated jets (\cite{komis07}).
Therefore, whether a model is based on large scale magnetic fields or not,
a confining external medium is required to form a collimated jet, and there is no
advantage to magnetic fields-based models.
The magnetic fields can further collimate the internal region, i.e.,
a self collimation (\cite{komis07}).

In Soker \& Regev (2003, hereafter SR03) and Soker \& Lasota (2004, hereafter SL04)
we reexamined the launching of jets from accretion disks by thermal pressure
following Torbett (1984; also Torbett \& Gilden 1992).
This paper describes the basic ingredients of the processes described in those papers,
as well as new ideas.

\section{Motivation}
There are several arguments that point to problems with models for launching jets
that are based only on large scale magnetic fields.

{\bf (1) Precessing jets.}
In several YSOs the jets precess on a time scale $\lesssim 100$~years (e.g. Barsony 2007).
A large scale magnetic field cannot change its symmetry axis on such a short time.

{\bf (2) A collimated jets in a planetary nebulae}
There is a highly collimate clumpy double-jet in the planetary nebulae Hen~2$-$90 (\cite{sahai2000}),
very similar in properties to jets from YSOs that form HH objects.
The source of the accreted mass in planetary nebulae is thought to be a companion star.
In such a system large scale magnetic fields are not expected.

{\bf (3) No jets in DQ Her (Intermediate polars) systems. }
Intermediate polars (DQ Her systems) are cataclysmic variables where the magnetic field of the
accreting WD is thought to truncate the accretion disk in its inner boundary.
This magnetic field geometry is the basis for some jet-launching models in YSOs
(e.g., Shu et al. 1991). However, no jets are observed in Intermediate polars.

{\bf (4) Thermal pressure. } Thermal pressure seems to be an
important ingredient even in MHD models for jet launching (e.g.,
Ferreira \& Casse 2004; Vlahakis et al. 2003; Vlahakis \& Konigl 2003).
In particular I note that in the exact solutions for
steady relativistic ideal MHD outflows found by Vlahakis \& Konigl
(2003; also Vlahakis et al. 2003) the initial acceleration phase
is by thermal pressure, and internal (thermal) energy is converted
to magnetic energy.

\section{Launching jets by thermal pressure gradients}

In YSOs the scheme was developed by Torbett (1984; also Torbett \& Gilden 1992),
and discussed in more detail by SR03. SL04 further discussed it, and include
accretion into white dwarfs (WDs) as well.
In this model, the accreted disk material is strongly shocked
due to large gradients of physical quantities in the boundary
layer, and then radiativelly cools on a time scale longer than
the ejection time from the disk.

The model assumes that hundreds of small blobs are
formed in the sheared boundary layer (section 2 of SR03).
The blobs occasionally collide with each other, and create
shocks which cause the shocked regions to expand in all directions.
If the shocked regions continue to expand out into the path of yet more
circulating blobs, stronger shocks may be created, as was proposed
by Pringle \& Savonije (1979) to explain the emission of X-rays
out of disk boundary layers in dwarf novae.
For the shocked blobs to expand, the radiative cooling time of
{\em individual blobs} must be longer than the adiabatic
expansion time of individual blobs.
SR03 demand also that the blobs be small, because the dissipation
time of disk material to form the strong shocks must be shorter than
the jet ejection time (eq. 24 of SR03).
SR03 find that the thermal acceleration mechanism works only when
the accretion rate in YSO accretion disks is large enough and
the $\alpha$ parameter of the disk small enough - otherwise the radiative
cooling time is too short and significant ejection does not take place.
SR03 term the strong shocks which are formed from the many
weakly shocked blobs, `spatiotemporally localized
(but not too small!) accretion shocks', or SPLASHes.
Such SPLASHes can be formed by the stochastic behavior of the magnetic fields
of the disk itself and of the central object, e.g., as in cases where the
inner boundary of the disk is truncated by the stellar magnetic field.

\begin{figure}  %
\vskip 0.1 cm
\resizebox{1.0\textwidth}{!}{\includegraphics{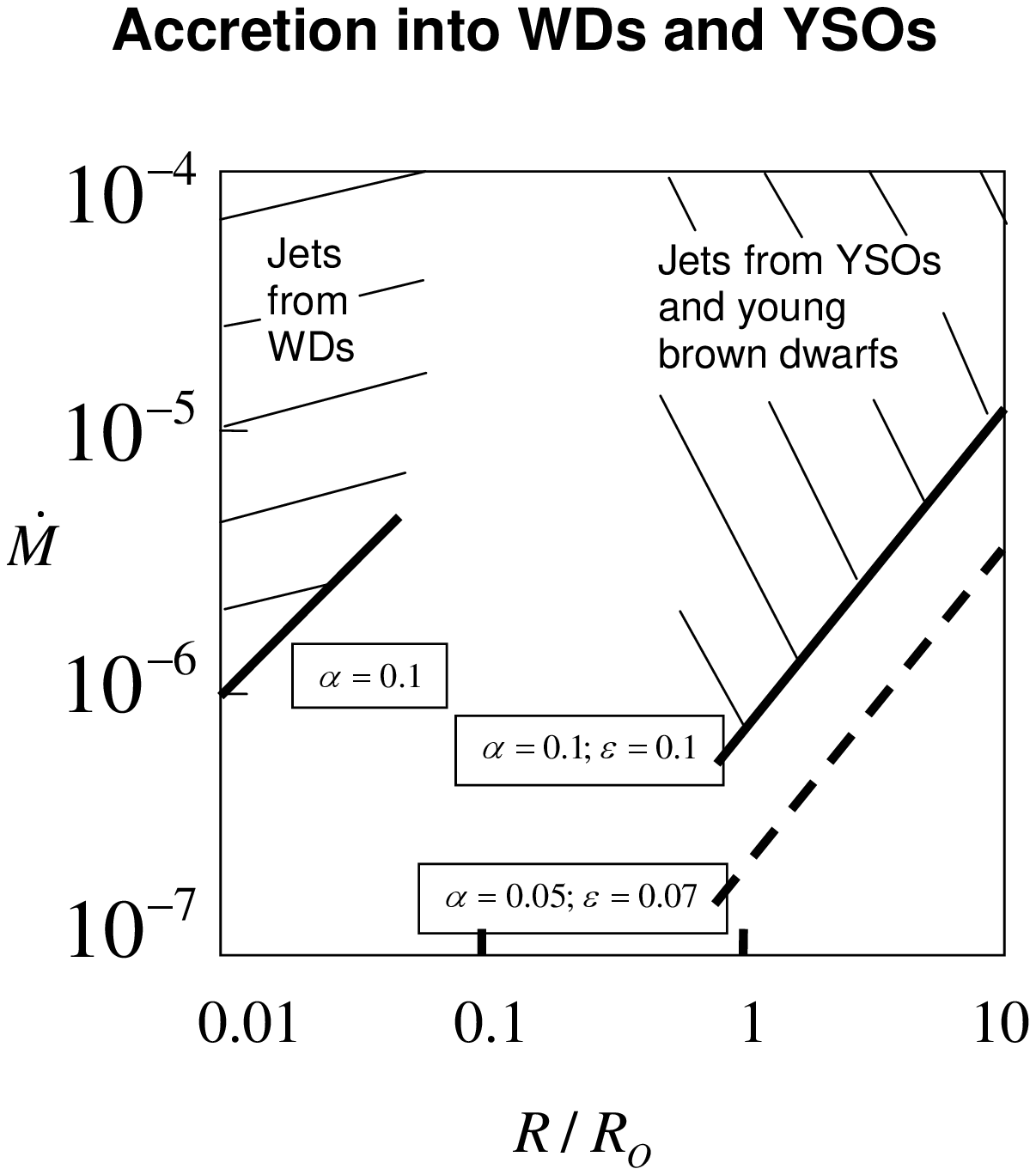}}
\vskip -7.0 cm  
\caption{The condition for the radiative cooling time to be longer than mass ejection time,
in the accretion rate (in $M_\odot \yr^{-1}$) versus jet launching radius (in $R_\odot$) plane.
Due to several uncertainties, the allowed mass accretions rate can be an order of magnitude lower,
but the general constraints are as shown.
The radiative cooling time of the postshock gas is dictated by
the photon diffusion time (SR03).
}
\label{ysojet}
\end{figure}

The model then has two conditions. The first condition is that the strongly
shocked gas in the boundary layer will cool slowly, such that the thermal
pressure will have enough time to accelerate the jet's material.
The radiative cooling is via photon diffusion.
The constraints translate to a condition on the mass accretion
rate to be above a minimum value, depending on the accreting objects,
e.g., a WD or a main sequence star.
The second condition, which in general is stringent, is that weakly shocked
blobs in the boundary layer will expand, and disturb the boundary layer in such a way that a strong
shock will develop.
This also leads to a minimum value for the mass accretion rate (SR03; SL04 eq. 12)
\begin{equation}
\dot M \gtrsim 4.2 \times 10^{-5}
\kappa^{-1}
\left( \frac{\alpha}{0.1} \right)
\left(\frac{R} {R_{\odot}} \right)
M_\odot \yr^{-1},
\label{acc01}
\end{equation}
where $R$ is the disk radius from where the jet is launched, $\alpha$ is the disk's
viscosity parameter, and $\kappa$ is the opacity.
Using this criterion SL04 find that the mass accretion rate above which jets could
be blown from accretion disks around YSOs, where $R \simeq R_\odot$ and
after substituting the opacity, is
\begin{equation}
\dot M_b({\rm YSOs}) \gtrsim
7 \times 10^{-7}  
\left( \frac{R}{R_\odot} \right)^{1.2}
\left( \frac{\epsilon}{0.1} \right)^{1.4}
\left( \frac{\alpha}{0.1} \right)
M_\odot \yr^{-1},
\label{accyso2}
\end{equation}
where $\epsilon = H/R$, and $H$ is the vertical disk's scale height.
This condition is drawn in the right hand side of Figure \ref{ysojet}
for two sets of ($\alpha; \epsilon$) parameters.

For WDs the post shock temperature is much higher than that YSOs, and the
opacity is $\kappa=0.4 \cm^2 \g^{-1}$.
This gives
\begin{equation}
\dot M_b ({\rm WD}) \gtrsim
10^{-6}
\left( \frac{R}{0.01R_\odot} \right)
\left( \frac{\alpha}{0.1} \right)
M_\odot \yr^{-1}.
\label{accwd2}
\end{equation}
This condition is drawn in the left hand side of Figure \ref{ysojet}.
SL04 noted that this limit is almost never satisfied in CVs (Warner 1995).
However, Retter (2004) suggested that this limit might be met during the
transition phase in novae, where a claim for a jet has been made.


\section{Jets from brown dwarfs (BDs)}

The thermal launching model presented here can work only when mass accretion rate is high enough.
I therefore turn to check the situation with brown dwarfs (BDs), where the claimed low
mass accretion rate was presented as an evidence against the thermal-launching model.

Whelan et al. (2005, 2007) argued for a BD jet similar to that observed in YSOs.
They use a forbidden line which forms at a critical density of
$2 \times 10^6 \cm^{-3}$, or $\rho_c \simeq 3 \times 10^{-18} \g \cm^{-3}$, achieved at
a distance $r_o$.
Assuming that the half opening angle of each jet is $\alpha \sim 10^\circ$, and
the jets speed is $v_j$, the outflow rate of the two jets combined is
\begin{equation}
2 \dot M_j =4 \pi \rho_c r_o^2 (1-\cos \alpha) v_j
\simeq 8 \times 10^{-10}
\left( \frac{r_o}{10 \AU} \right)^2
\left( \frac{\alpha}{10^\circ} \right)^2
\left( \frac{v_j}{40 \km \s^{-1}} \right)  M_\odot \yr^{-1}.
\label{dotm1}
\end{equation}
I substitute numbers as given by Whelan et al. (2007, Table 1) for two systems.
For 2MASS1207-3932 I take for the observed values $v_j=8 \km \s^{-1}$ and $r_o=4 \AU$,  while
for $\rho-$Oph~102 I take $v_j=40 \km \s^{-1}$ and $r_o=10 \AU$.
The unknown inclination implies that both the distance and velocity are larger than the
observed values, and the mass outflow rate should be multiply by a number $>2.6$;
I multiply it by 2.6.

For 2MASS1207-3932 I find $\dot M(2MASS1207-3932) \simeq 6.6 \times 10^{-11} M_\odot \yr^{-1}$. From
Figure 2 of \cite{Whelan2007} it seems that the blue-shifted outflow has a large
opening angle (or large covering factor), and I expect that for this case $\alpha >10^\circ$.
Over all, the mass loss rate is $\dot M(2MASS1207-3932) \gtrsim 10^{-10} M_\odot \yr^{-1}$.
This is an order of magnitude larger than the accretion rate given in Table 1 of \cite{Whelan2007}.
For $\rho-$Oph~102 I find  $\dot M(\rho-{\rm Oph}~102) \simeq 2 \times 10^{-9} M_\odot \yr^{-1}$.
This mass outflow rate is 60\% higher than the mass accretion rate given in Table 1 of
\cite{Whelan2007}.
Jets with opening angle much smaller that $10^\circ$ are not likely close to their source
in YSOs.
Considering that mass outflow rates in YSOs are $0.01-0.1$ times the mass accretion rate,
I conclude that there is inconsistency in the data given by Whelan et al. (2005, 2007)
for outflow from BDs.
The resolution to this problem can be one of the following.
\newline
(1) The accretion rate is indeed very low ($\sim 10^{-11} M_\odot \yr^{-1}$),
and the mass outflow rate is much smaller
than what I estimated above. In that case, the outflow rate ($\sim 10^{-12} M_\odot \yr^{-1}$)
can be accounted for by a BD stellar-type wind, with no need for jets.
\newline
(2) The mass accretion rate is much higher than that given by \cite{Whelan2007}
and Mohanty et al. (2005), i.e.,  it is  $\dot M_{\rm acc} \simeq 10^{-8} M_\odot \yr^{-1}$.
 In that case jets can be launched according to the model presented here.
 A higher mass accretion rate is suggested also by the young age of accreting BDs, which
 is similar to that of YSOs (Mohanty et la. 2005).
 According to Mohanty et al. (2005) the number of accreting WDs declines substantially
 by the age of $\sim 10^7 \yr$. For a BD mass of $M_{\rm BD} > 0.01 M_\odot$ the implied
 average accretion rate should be $ \gtrsim 10^{-9} M_\odot \yr^{-1}$.
Short accretion phase of YSOs is suggested also by the work of Lucas Cieza et al (2007).
\newline
(3) There are very large variations on short time scales of the mass accretion rate,
as suggested by \cite{Scholz2006}. Jets are launched then only during the very high
mass accretion phases. On average, the mass accretion rate can stay low.
The outflow is composed of many small clumps, and the {\it average} mass outflow rate is
much smaller than what I calculated above.

I predict that future studies will show that the mass accretion rate into BD that launch jets
(and not stellar-type winds) is $\dot M_{\rm acc}(BDs) \gtrsim 10^{-9} 10^{-8} M_\odot \yr^{-1}$, i.e.,
larger than current studies show (e.g., Mohanty et la. 2005).
Such accretion rate with low enough value of the disk-viscosity parameter $\alpha$
(see equation 3.2), is compatible with the model presented here.



\section{A DOUBLE-STAGE ACCELERATION SCENARIO}
\begin{figure}  %
\vskip -1. cm
\resizebox{0.99 \textwidth}{!}{\includegraphics{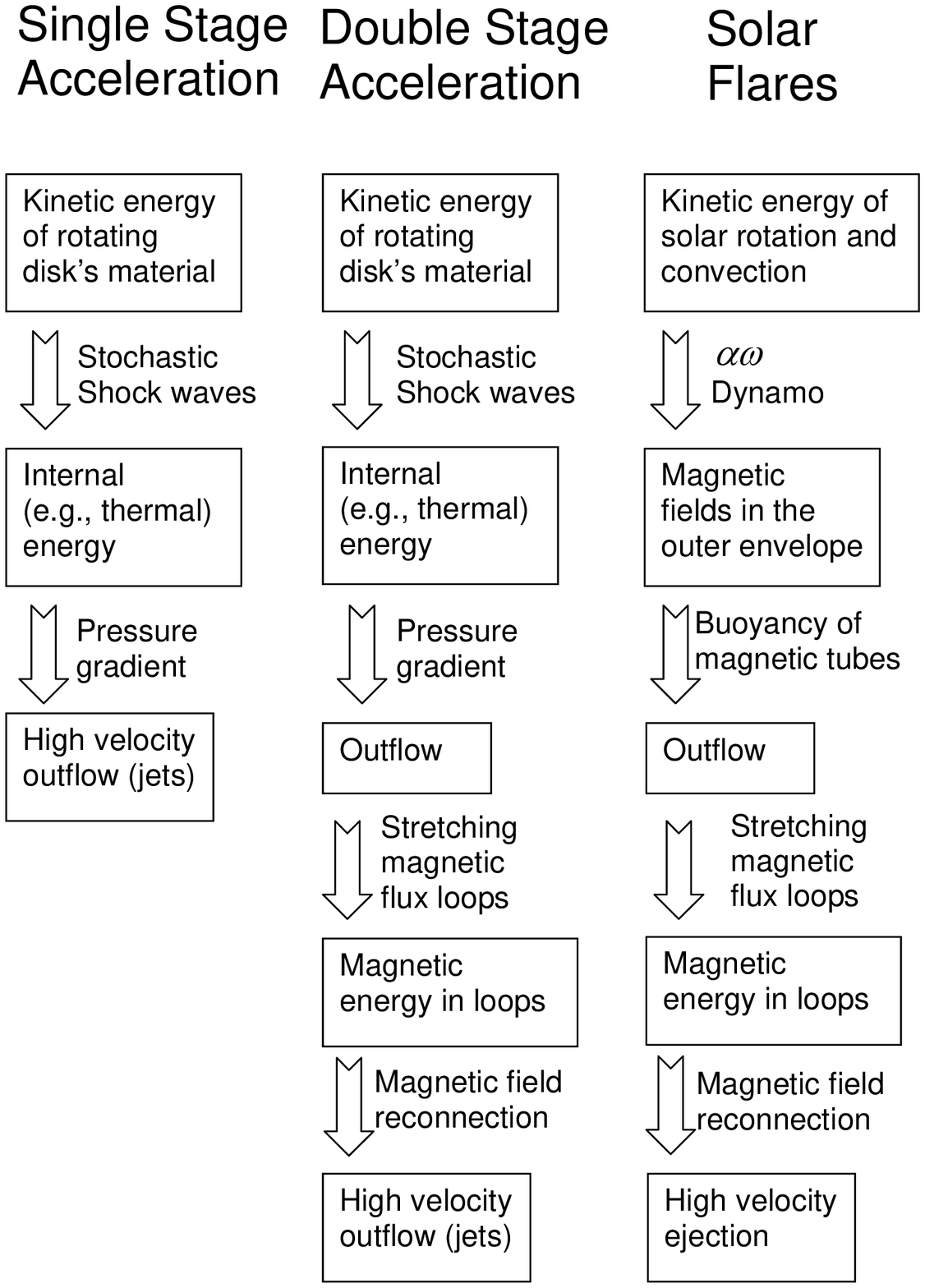}}
\vskip -3.0 cm  
\caption{ The energy transformations in the single-stage acceleration
scenario and in the double-stage acceleration scenario.
}
\label{ennergf}
\end{figure}

I propose the possibility that thermal energy is transferred to
magnetic energy, via kinetic energy of the outflowing gas, and a second stage of
acceleration takes place wherein magnetic field reconnection accelerates the gas to
velocities much above the escape velocity, as in solar flares.
The energy cycle is as depicted in Figure \ref{ennergf}.

The idea that magnetic energy can serve as an intermediate stage is not new.
In launching jets from BHs and NSs the dissipated bulk kinetic energy
might be channelled to magnetic energy, and then to kinetic energy.
In the exact solutions for steady, relativistic, ideal MHD outflows found by
Vlahakis \& K\"onigl (2003; also Vlahakis et al. 2003) the initial acceleration
phase is by thermal pressure, and internal (thermal) energy is converted to
magnetic energy.
I consider a non-steady state outflow, based on flare-like ejection, as in solar
coronal mass ejection (CME), rather than a steady state outflow.
Direct build-up of magnetic fields, without the intermediate stage of
thermal energy, is discussed by Machida \& Matsumoto (2003, their sec. 4)
who show how in the plunging region of the accretion disk around BHs, where gas
falls to the BH and no stable orbits are possible, the gravitational energy of
the accreting gas is converted to magnetic energy.

The conversion of accretion energy (kinetic energy of the rotating disk or gravitational energy)
to thermal energy, then to magnetic field, and then to the kinetic energy of the ejected wind
will be studied in a future paper.

\begin{acknowledgments}
I thank Oded Regev for many useful comments.
This research was supported by the Asher Fund for Space Research at the
Technion.
\end{acknowledgments}

\end{document}